\documentclass[%
reprint,
nofootinbib,
mathtools,
amsmath,
amssymb,
prd,
longbibliography,
superscriptaddress]{revtex4-2}

\usepackage{booktabs}
\usepackage{mathrsfs}
\usepackage{graphicx}
\usepackage{dcolumn}
\usepackage{bm}
\usepackage{lipsum}
\usepackage{graphics}
\usepackage{bbm}
\usepackage[dvipsnames]{xcolor}
\usepackage{color}
\usepackage{microtype}
\usepackage{IEEEtrantools}
\usepackage{mathrsfs}
\usepackage{amsthm}
\usepackage{enumitem}
\usepackage[normalem]{ulem}
\usepackage[pdftex,breaklinks,colorlinks,
linkcolor=Blue,
citecolor=teal,
anchorcolor=red,
urlcolor=cyan]{hyperref}
\usepackage{orcidlink}

\newcommand{\scri}{\mathcal{I}}
\def\nn{\nonumber}
\begin{document}

\title{A proof of conservation laws in gravitational scattering: \\ 
 tails and breaking of peeling}

\author{Geoffrey Comp\`ere\,\orcidlink{0000-0002-1977-3295}}
\author{S\'ebastien Robert\,\orcidlink{0009-0009-8592-1768}}
\affiliation{%
 Universit\'e Libre de Bruxelles, BLU-ULB Brussels Laboratory of the Universe, C.P. 231, B-1050 Bruxelles, Belgium
}%

\date{\today}

\begin{abstract}

We propose a definition of asymptotically flat spacetimes that is consistent with both null infinities and compatible with known properties of gravitational scattering, incoming and outgoing radiation, and interactions with matter. For this class of spacetimes, we prove three antipodal matching conditions at spatial infinity: one for the so-called dual mass aspect, one for the leading tail of the shear, and one that non-trivially relates the peeling properties of the spacetime at both null infinities to the leading tail and mass aspect at spatial infinity. Furthermore, we reformulate these identities as conservation and flux balance laws  defined on the boundary hyperboloid at spatial infinity. 
\end{abstract}

\maketitle

\emph{Introduction.}
There is currently no universally accepted definition of spacetimes without a cosmological constant capable of describing the full range of non-linear gravitational dynamics including binary black hole mergers, $n$-body scattering, and interactions with matter. This fundamental question lies at the core of recent progresses in the infrared properties of gravity \cite{Strominger:2017zoo}, in the formulation of non-linear gravitational dynamics \cite{Christodoulou:2008nj} and in the reformulation of gravity as a holographic theory \cite{Pasterski:2021raf}. 

Though physically relevant to describe a restricted class of isolated gravitational systems, the ``asymptotically simple'' class of spacetimes \cite{Penrose1962,Penrose:1964ge,1977asst.conf....1G,Frauendiener:2000mk} obeying the peeling property of the Weyl tensor \cite{sachs1961gravitational,Penrose1962,10.1098/rspa.1965.0058} is \emph{not} sufficiently general to encompass spacetimes with both incoming and outgoing radiation \cite{PhysRevD.19.3495,Damour:1985cm,1985FoPh...15..605W,christodoulou2002global,Kehrberger:2021uvf}, nor $n$-body scattering \cite{christodoulou2002global,Kehrberger:2021vhp,Sahoo:2021ctw,Laddha:2018myi}. 


A proposal for an asymptotic framework compatible with scattering and radiation and which admits the BMS group \cite{bms2,bondi1962gravitational} as asymptotic symmetry group has been proposed in \cite{Compere:2023qoa} following earlier work \cite{1972JMP....13..956G,Ashtekar:1978zz,beig1982einstein,Friedrich:2002ru,Henneaux:2019yax,Prabhu:2021cgk,chakraborty2022supertranslations,Capone:2022gme}. However, it is incomplete, as it does not include spacetimes breaking peeling \cite{PhysRevD.19.3495,Damour:1985cm,ChristodoulouKlainerman+1994,Friedrich:2017cjg,Saha:2019tub,Kehrberger:2021uvf,Kehrberger:2021vhp,Kehrberger:2021azo,Masaood:2022bvi,Kehrberger:2024clh,Kehrberger:2024aak,Schneider:2025tek,1985FoPh...15..605W}. At future null infinity, the polyhomogeneous Bondi expansion has been formulated   \cite{1985FoPh...15..605W,chrusciel1995gravitational,Kroon:1998tu,ValienteKroon:2002gb,Hintz:2017xxu} and has been shown to also admit the BMS group as asymptotic symmetry group~\cite{Geiller:2024ryw}. 
However, no consistent linking of the metric between both null infinities across spatial infinity up to subleading order in the respective expansions has been derived in the non-linear theory, which prevents the formulation of a coherent scattering theory  (for progress in an alternative geometric approach, see  \cite{Ashtekar:2023wfn,Ashtekar:2023zul,Borthwick:2024skd,Marajh:2025yhg}).

In this Letter, we set a framework based on polyhomogeneous Bondi expansions at both null infinities that allows to prove exact maps between features of the asymptotic radiation and the peeling behavior at future and past null infinities. We work in the non-linear theory including matter interactions and both incoming and outgoing radiation with tails compatible with gravitational scattering. We also allow for a leading magnetic component of the shear near spatial infinity (but disallow NUT charges), which admits potentially physical features \cite{Satishchandran:2019pyc,Godazgar_2019,Kol_2019}. We warn the reader that we do not prove that our assumed asymptotic behavior at past and future null infinity originate from classes of initial data; this question is addressed in complementary analyses, see e.g. \cite{ValienteKroon:2002gb,Hintz:2017xxu,Kehrberger:2024clh,Fang:2024ivh,Marajh:2025yhg}.

In what follows, we  describe our detailed hypotheses and derive the three identities \eqref{MAIN}, which constitute   our main result. We further reformulate these identities as conservation and flux balance laws at spatial infinity. Our results clarify the conditions under which peeling breaks in the class of systems described by our hypotheses. The technical expressions and proofs are relegated to the Supplementary Material and crucially rely upon harmonic analysis on three-dimensional de Sitter spacetime \cite{Compere:2025bnf}.

\emph{Hypotheses.}
At future null infinity $\scri^+$, we work in Bondi gauge where the metric takes the form 
\cite{bondi1962gravitational,sachs1961gravitational,Sachs1962BMS,bondi1960gravitational,sachs1962gravitational} 
\begin{align}
    ds^2=&\frac{V e^{2\beta}}{r} du^2 - 2 e^{2\beta}du dr \nn\\&+ r^2 h_{AB} (dx^A-U^Adu)(dx^B-U^Bdu).
\label{BondiSachsMetric}
\end{align}
We consider a polyhomogeneous expansion of the form 
\begin{subequations}\label{BondihExp}
\begin{align}
    V&:=-r+2m+{o}(r^{0})\label{BondiVExp},\\
    \beta&:=\frac{\beta_1}{r}+\frac{\beta_2}{r^2}+o(r^{-2})\label{BondiBExp},\\
    U^A&:=\frac{U_2^A}{r^2}+\frac{\log r \, U_{3,1}^{A}+U_3^A}{r^3}+o(r^{-3})\label{BondiUExp},\\
    h_{AB}&:=\gamma_{AB}\sqrt{1+\frac{\mathcal C^{CD}\mathcal C_{CD}}{2r^2}}+ \frac{\mathcal C_{AB}}{r}, \\
    \mathcal C_{AB} &:= C_{AB}+\frac{1}{r}D_{AB}+o(r^{-1}).  
\end{align}    
\end{subequations}
Capital Latin indices are raised and lowered with the metric $\gamma_{AB}$; $\nabla_A$ denotes its Christoffel connection, $\epsilon_{AB}$ its Levi-Civita symbol and $\int_{S^2}d^2\Omega=\int_{S^2}d^2 x \sqrt{\gamma} =4\pi$ its unit measure. The Bondi fields $C_{AB}$ and $D_{AB}$ are traceless with respect to $\gamma_{AB}$. The superrotation frame is fixed by choosing $\gamma_{AB}$ as the sphere metric. The news $N_{AB}:= \partial_u C_{AB}$ transforms homogeneously under the BMS group  \cite{Barnich:2009se,Compere:2018ylh,Campiglia:2020qvc,rignonbret2024centerlessbmschargealgebra}. We assume that all Bondi aspects are smooth over the sphere, which excludes NUT charges. The polyhomogeneous expansion \eqref{BondihExp} was thoroughly studied in \cite{Geiller:2024ryw} following earlier work \cite{1985FoPh...15..605W,chrusciel1995gravitational,Geiller:2022vto}. Using the Bondi tetrad $(l,n,m,\bar m)$, the tensor $D_{AB}$ corresponds to the lack of peeling in $\Psi_0 = r^{-4}D_{AB}m^A m^B+o(r^{-4})$ and $\Psi_1=\nabla^B D_{AB}m^A \log r\,  r^{-4}+O(r^{-4})$ for vacuum solutions of Einstein's equations \cite{Geiller:2024ryw}. 

The metric is defined at past null infinity $\scri^-$ from the formally identical expansion as at $\scri^+$ with $u$ replaced by $-v$.  We introduce the notation $u^+ :=  u$ and $u^- := -v$ in order to write expansions at $\scri^+$ and $\scri^-$ with a notation that encompasses both cases. In order to distinguish the coefficients of the Bondi-Sachs expansion between $\scri^+$ and $\scri^-$ we promote all Bondi-Sachs quantities with an additional $+$ superscript, e.g. $C^+_{AB} := C_{AB}$ and we write all Bondi-Sachs quantities at $\scri^-$ with a $-$ superscript. In particular, the news at $\scri^-$ is defined as $N^-_{AB} := -\partial_v C_{AB}$. The limit $u \to -\infty$ of $\scri^+$ is denoted as $\scri^+_-$ while the limit $v \to +\infty$ of $\scri^-$ is denoted as $\scri^-_+$. The locations $\scri^+_-$ and $\scri^-_+$ will be referred to as the ``corners'' of the spacetime at spatial infinity defined at $u^\pm \to - \infty$. 

Since we aim to describe the most general scattering processes allowed by the classical soft theorems \cite{Saha:2019tub,Choi:2024ajz}, including long-range interactions, we assume that the shear $C^\pm_{AB}$ at both $\scri^+$ and $\scri^-$ behaves in the approach to the corners $u^\pm \rightarrow -\infty$ as follows 
\begin{align}
    C_{AB}^\pm & = C_{AB}^{\pm (0)}(x^A)-\frac{C_{AB}^{\pm (1)}(x^A)}{u^\pm}+o(\frac{1}{u^\pm})\label{CABasymptoticsscri}.
\end{align}

The terms $C_{AB}^{\pm (1)}(x^A)$ are the leading order tails at spatial infinity. We do not impose the vanishing of the leading-order magnetic part of the shear $C^{\pm(0)(B)}_{AB}$, which would correspond to setting $\epsilon_{CA}\nabla^A\nabla^B C^{\pm(0)\;C}_B = 0$. Allowing $C^{\pm(0)(B)}_{AB}\neq 0$ allows solutions whose linearization around Minkowski was  described in \cite{Satishchandran:2019pyc}, which produces a magnetic displacement memory effect. 

The stress-energy tensor compatible with Eq. \eqref{BondihExp} is given by Eqs. \eqref{Tuu}-\eqref{TAB} in Appendix \ref{sec:EinsteinNullInfty}. This generalizes the expansion considered in \cite{Flanagan:2015pxa}. For simplicity, we assume that the stress-energy tensors on $\scri^\pm$ are compactly supported and do not appear at $i^0$ at the order considered in this paper. We could relax this assumption to allow for suitably defined boundary conditions in the limit to the corners but this analysis is not presented here.

 

The class of spacetimes considered allows (both incoming and outgoing) gravitational radiation, massive and massless particles, i.e. generic gravitational scattering. In that case, the leading order magnetic part of the shear vanishes \cite{Sahoo:2021ctw}. This can be derived from Einstein's equations at timelike infinity after substituting the stress-energy tensor of multipolar point-particles \cite{Boschetti:2025tru}.  

\emph{Results.} Under this set of assumptions, the following quantities defined either at $\scri^+$ or  $\scri^-$
\begin{subequations}
\begin{align}
    \tilde{\mathcal{M}}^\pm&:=\frac{1}{4} \nabla_C \nabla^E\epsilon^{CD} C^\pm_{DE}- \frac{1}{8} \epsilon^{CD} C^\pm_{DE} N^{\pm \,EC}\\
 \mathcal{C}^{\pm(1)}_{AB} &:= (u^\pm)^2 N^\pm_{AB}+(-E\pm P_in_i)C_{AB}^{\pm(B)},\\ 
   \mathcal D_{A}^\pm &:= \nabla^B D^\pm_{AB} \pm 6 P_i (m^\pm \nabla_A n_i +\tilde{\mathcal{M}}^{\pm}\epsilon_{AB}\nabla^B n_i)\nn\\&\qquad - 2 (E \mp P_i n_i) (\nabla_A m^\pm+\epsilon_{AB}\nabla^B\tilde{\mathcal{M}}^{\pm})\nn\\&\qquad -\frac{(u^\pm)^2}{2} \nabla_C \nabla_{\langle A} \nabla_{B\rangle} N^{\pm BC} ,\label{cDA}
\end{align}    
\end{subequations}
admit a finite limit at the corners, $\text{lim}_{u^\pm \rightarrow -\infty}\tilde{\mathcal{M}}^\pm=\frac{1}{4} \nabla_C \nabla^E\epsilon^{CD} C_{DE}^{\pm(0)}:=\tilde{\mathcal{M}}^{\pm(0)}$, $\text{lim}_{u^\pm \rightarrow -\infty}    \mathcal{C}^{\pm(1)}_{AB}  = C^{\pm (1)}_{AB}+(-E\pm P_in_i)C_{AB}^{\pm(B)(0)}:=\mathcal{C}^{\pm(1)(0)}_{AB}$ and $\text{lim}_{u^\pm \rightarrow -\infty} \mathcal D_{A}^\pm := \mathcal D_{A}^{\pm (0)}$. Here $E$ and $P_i$ are the total energy and momentum of the spacetime defined at spatial infinity and $\tilde{\mathcal{M}}$ is the dual covariant mass aspect, which is the leading $r^{-3}$ coefficient of the imaginary part of the Weyl scalar $\Psi_2$ in the Bondi tetrad  \cite{Freidel:2021qpz,Geiller:2024ryw}. The quantities  $\tilde{\mathcal{M}}^{\pm (0)}$ correspond to the dual supertranslation charges which can be non-vanishing only for $\ell \geq 2$ spherical harmonics \cite{Kol_2019,Godazgar_2019N} (as the $\ell=0,1$ modes are excluded by the absence of NUT charges). 

In this Letter, we prove that these corner quantities are boundary values of charges defined at spatial infinity $i^0$, i.e. integrals over the sphere of tensor quantities on the boundary hyperboloid $dS_3$. We demonstrate that the boundary values $\tilde{\mathcal{M}}^{\pm(0)}$, $C_{AB}^{\pm (1)}$ and $\mathcal D_{A}^{\pm (0)}$ admit the following antipodal matching relationships between the corners:
\begin{subequations}\label{MAIN}
\begin{align}
&\Upsilon^* \tilde{\mathcal{M}}^{+(0)} = -\tilde{\mathcal{M}}^{-(0)},\label{idCABB}\\
&\Upsilon^* \nabla^B C_{AB}^{+(1)}= \nabla^B C_{AB}^{-(1)}+2\Upsilon^*\mathcal S_A [\mathcal D_B^{-(0)}],\label{idCAB}\\&
   \Upsilon^* \mathcal D_A^{+(0)}  = - \mathcal D_A^{-(0)}, \label{idDA} 
\end{align}    
\end{subequations}
where $\Upsilon^*$ is the action of the parity flip over the sphere and $\mathcal  S_A[\mathcal D_B^{-(0)}](x^A)$ will be defined after Eq. \eqref{map1}.  

The first identity proves the conjectured antipodal matching condition for the dual mass aspect  
\cite{Godazgar_2019,Kol_2019}. It is equivalent to $\Upsilon^* C_{AB}^{+(0)(B)}=  C_{AB}^{-(0)(B)}$. It was obtained  under a different set of hypotheses in linearized gravity in \cite{Masaood:2022bvi}. The second and third identities are novel. The third identity could be called the leading law of peeling, since it determines whether or not peeling holds at leading order in the radial expansion at $\scri^+$ in terms of properties at $\scri^-$. It was independently and nearly simultaneously derived in \cite{Boschetti:2026gfd}. 

\emph{Derivation.}
In order to derive Eqs. \eqref{idCAB}-\eqref{idDA}, one first needs to solve Einstein's equations at $\scri^\pm$ to subleading order in the Bondi-Sachs expansion. The components $G_{r\mu}=8 \pi T_{r\mu}$ lead to algebraic constraints while the remaining equations lead to three evolution equations for a basis of Bondi aspects up to subleading order, which we choose as the set ($m^\pm,\mathcal{P}^\pm_A,D^\pm_{AB}$), see Appendix \ref{sec:EinsteinNullInfty}. In the presence of sufficiently generic matter, peeling generically does not hold as $D^\pm_{AB}$ is sourced by the asymptotic stress-energy tensor, as demonstrated in Eq. \eqref{FluxBalancedLawsD}.

Then, using assumption \eqref{CABasymptoticsscri} the fields $m^\pm$ and $\mathcal{P}^\pm_A$ must behave as follows in the approach to the corners as $u^\pm \rightarrow -\infty$:
\begin{subequations}\label{falloffIplus}
\begin{align}
m^\pm & = m^{\pm (0)} -\frac{m^{\pm (1)}}{u^\pm}+o\big( \frac{1}{u^\pm}\big)\label{mapectasymptotics}, \\
\mathcal{P}^\pm_A & =- N_A^{\pm(-1)} u^\pm+N_A^{\pm (log)}\log(-u^\pm)+ N_A^{\pm (0)}+o(1),
\end{align}
\end{subequations}
where 
\begin{align}
m^{\pm (1)} &:=\frac{1}{4} \nabla^A \nabla^B C^{\pm (1)}_{AB},\label{EvolutionConstrainm1}\\
N_A^{\pm (-1)}&:= -\nabla_A m^{\pm (0)}-\frac{1}{4} \epsilon_{AB} \nabla^B \epsilon_{CD} \nabla^D \nabla^E C^{\pm (0)\;C}_{E},
\end{align}
\begin{align}
    N_A^{\pm (log)} &:=-\frac{1}{2} \nabla_C \nabla_{\langle A} \nabla_{B\rangle} C^{\pm (1)\;BC}\label{NALog}. 
\end{align}
It is then proven that $D_{AB}^{\pm}$, $\mathcal C^{\pm (1)}_{AB}$ and $\mathcal D^\pm_A$ admit a finite limit as $u^\pm \rightarrow -\infty$ where we identify the finite  $\mathcal D_A^{\pm (0)}$ as
\begin{align}
\mathcal D_A^{\pm (0)} & =\nabla^B D_{AB}^{\pm(0)} - \frac{1}{2} \nabla_C \nabla_{\langle A} \nabla_{B\rangle} C^{\pm (1) BC} \nn\\&\quad \pm 6 P_i (m^{\pm (0)}\nabla_A n_i +\tilde{\mathcal{M}}^{\pm(0)}\epsilon_{AB}\nabla^B n_i)\nn\\&\quad- 2 (E\mp P_i n_i) (\nabla_A m^{\pm (0)}+\epsilon_{AB}\nabla^B\tilde{\mathcal{M}}^{\pm(0)}) ,\label{defDA}
\end{align}
with $D_{AB}^{\pm(0)}:=\lim_{u^\pm\to-\infty}D_{AB}^{\pm}$.

Beig-Schmidt coordinates \cite{beig1982einstein,beig1984integration} are denoted as $(\rho, \phi^a)$, $a=1,2,3$, with $\phi^a = (\tau, x^A)$. Spatial infinity is defined in the limit $\rho \rightarrow \infty$ and the corners are matched in the subsequent limit $\tau \rightarrow \pm \infty$  \cite{Compere:2023qoa}. For the class of metrics \eqref{BondiSachsMetric}-\eqref{BondihExp} and under the assumptions for the shear \eqref{CABasymptoticsscri}, we can perform the change of coordinates from Bondi to Beig-Schmidt coordinates in the vicinity of both corners, see Appendix \ref{sec:Matching}. This change of coordinates at both corners therefore exists up to subleading order in the $\rho \rightarrow\infty $ expansion while keeping $k^{(B)}_{ab}$ non-vanishing at leading order and allowing for a violation of peeling. This significantly improves the scope of earlier results \cite{Capone:2022gme,Compere:2023qoa}. The Beig-Schmidt expansion obtained in the limits $\tau \to \pm\infty$ reads as
\begin{align}
     & ds^2= \left(1+\frac{2\sigma}{\rho}+\frac{\sigma^2}{\rho^2}+o(\rho^{-2})\right)d\rho^2 + o(\rho^{-2}) \rho d\rho d\phi^a\nn \\&+\rho^2 \bigg(\! q_{ab} \!+ \! \frac{k_{ab}-2\sigma q_{ab}}{\rho} \!+\!\frac{\log\rho}{\rho^2}   i_{ab}\!+\!\frac{j_{ab}}{\rho^2}\!+\!o(\rho^{-2})\!\bigg)d\phi^ad\phi^b.   \label{BSExpansion}
\end{align}
The leading order metric is Minkowski in hyperboloidal coordinates with $q_{ab}$ the unit metric over $dS_3$. In what follows, lowercase Latin indices are raised and lowered with this metric, $\mathcal D_a$ is its metric-compatible covariant derivative, $\sqrt{-q}$ its measure and $n_a := -\partial_a \tau$. We will use the time-reversal antipodal map $\Upsilon_{\mathcal H}$ defined as the combination of parity $\Upsilon$ and time reversal $\tau \mapsto - \tau$. The field $k_{ab}$ has a vanishing trace, $k := k_{ab}q^{ab}=0$, and admits an electric and magnetic part $k^{(B)}_{ab}$ as $k_{ab}= -2 (\mathcal D_a \mathcal D_b + q_{ab})\Phi+k^{(B)}_{ab}$. The trace condition $k=0$ is therefore equivalent to $(\mathcal D_c \mathcal D^c + 3)\Phi = 0$. 

Given the asymptotic data at $\tau \to -\infty$ obtained from the matching at $\mathcal I_+^-$ and the fact that all Beig-Schmidt fields obey wave equations on $dS_3$ with well-defined evolution \cite{Compere:2025bnf}, we find that the Beig-Schmidt expansion \eqref{BSExpansion} remains consistent for all $\tau$. In particular, no other branch of the Beig-Schmidt expansion \eqref{BSExpansion} exists given our hypotheses (such as e.g. a $\log^2\rho$ branch), as any such branch obeys a wave equation with vanishing initial data as a consequence of the matching with $\mathcal I_+^-$ and therefore vanishes for all $\tau$.

The residual coordinate transformations of Beig-Schmidt gauge are detailed in Appendix \ref{sec:residualTrans}. Fields appearing in the expansion are defined over $dS_3$ and satisfy a set of equations after imposing Einstein's equations, see  Appendix \ref{sec:EIN} and \cite{Compere:2011ve} for a detailed derivation.

The result of the map is as follows. The field $\sigma$ converges at both corners up to a logarithmic translation as in \cite{Compere:2023qoa}. It can be decomposed in terms of its even $\bar\sigma$ and odd $\eta$ parts with respect to the antipodal map on the hyperboloid $\Upsilon_H$ as 
$\sigma = \bar \sigma + \eta$. The odd part $\eta$ only contains spherical $\ell=0,1$ harmonics and is shifted by logarithmic translations. As in \cite{Compere:2023qoa} the field $\sigma$ vanishing at $\tau \rightarrow \pm \infty$ has $\eta=\pm H_0$ with $H_0 := -2 E \sinh \tau +2 P^i n_i \cosh\tau$.  Since $\mathcal D_a \mathcal D_b + q_{ab}$ cancels the $\ell=0,1$ odd scalar harmonics and the matching conditions exclude the $\ell=0,1$ even scalar harmonics, we assume without loss of generality that $\Phi$ does not contain them. At leading order, it was found \cite{Troessaert:2017jcm,Prabhu:2019fsp,Magdy:2021rmi,Compere:2023qoa} that the leading behavior of $\sigma$ corresponds to the mass aspect and satisfies even antipodal matching condition, $m^{+(0)}=\Upsilon ^* m^{-(0)}$, while the leading behavior of $\Phi$ matches with the leading order electric part of the shear and satisfies odd antipodal matching conditions.

Since the Beig-Schmidt fields obey linear equations at subleading order, we can consider the leading order magnetic part of the shear separately. Since $k_{ab}$ satisfies the homogeneous equation $(\Box-3)k_{ab}$ and is symmetric divergence-free and traceless (SDT), see Appendix \ref{sec:EIN}, we can use the results for $n=-1$ SDT tensors in \cite{Compere:2025bnf}. For $n=-1$ magnetic parity SDT tensors, there exist two kinds of solutions: $p$-parity solutions which are even under the antipodal map $\Upsilon_\mathcal{H}$ and behave as $e^{\vert\tau\vert}$ for $\tau \rightarrow \pm\infty$ and the $q$-parity solutions which are odd under the antipodal map $\Upsilon_\mathcal{H}$ and behave as $\tau e^{\vert\tau\vert}$ for $\tau \rightarrow \pm\infty$. Using \eqref{kabscri+-}, we find that $k^{(B)}_{ab}$ is even on the hyperboloid and matches with the magnetic piece of $C_{AB}^{(0)}$. Therefore, we have derived  Eq. \eqref{idCABB}. The sign is opposite to the antipodal matching condition of the electric piece of $C_{AB}^{(0)}$. 

At subleading order, the fields of interest $i_{ab}$ and $j_{ab}$ transform non-trivially under logarithmic translations and supertranslations, see \cite{Compere:2023qoa} and Appendix \ref{sec:residualTrans}. 
Since the logarithmic translation gauge differs between the corners, it is more convenient to build invariant tensors under those transformations. It is also more practical to work with supertranslation invariant fields which can be chosen to be SDT tensors in the case of sub-subleading fields. Using the transformation laws \eqref{sigmaTransH}-\eqref{deltajab} and \eqref{BSEquationOrder1}-\eqref{NLsigmakab}, we built two invariant SDT tensors $I_{ab}$, $J_{ab}$, respectively from $i_{ab}$ and $j_{ab}$, see Eqs. \eqref{defIab}-\eqref{defJab}. These two invariant SDT tensors satisfy the equations 
\begin{align}
I^{a}_a = \mathcal{D}^b I_{ab} &= 0,\qquad 
     (\Box - 2)I_{ab} = 0,\label{eq:99b}\\
J_{a}^{a} = \mathcal{D}^b J_{ab} &= 0,\qquad 
     (\Box - 2)J_{ab} = S_{ab}\label{BoxJab},
\end{align}
where the asymptotic behavior of $S_{ab}$ as $\tau \rightarrow \pm \infty $ is given in Eq. \eqref{SabASY}. The source is $2I_{ab}$ plus terms that are even ($q$-parity) under $\Upsilon_{\mathcal H}$. 

We now compute the asymptotic behavior of $I_{ab}$ and $J_{ab}$ at $i^0_+=\scri^+_-$. This is achieved by injecting Eqs. \eqref{sigmascri+-}-\eqref{jABscri+-} into the definitions \eqref{defIab}-\eqref{defJab} with $\eta=+H_0=-2E \sinh\tau+2P_in_i$. The asymptotic behavior at $i^0_-=\scri^-_+$ is obtained by replacing $\tau\to-\tau$ and using the logarithmic translation gauge $\eta=-H_0$. Using Eq. \eqref{NALog} combined with Eq. \eqref{defDA}, we find  
\begin{align}
I_{\tau A} & = \pm 4 e^{-2 \vert\tau\vert} \mathcal{D}_A^{\pm(0)}+o\left(e^{-3 \vert\tau\vert}\right) ,\label{IetaabtauA+}\\
J_{AB}  &=\frac{1}{2}\big(C^{\pm(1)}_{AB}+(-E\pm P_in_i)C_{AB}^{\pm(B)(0)}\big) e^{2\vert\tau\vert}+o\left(e^{2 \vert\tau\vert}\right)\label{JetaABscri+-} . 
\end{align}

The tensor $I_{ab}$ is a  SDT tensor that satisfies the homogeneous equation $n=0$ as studied in \cite{Compere:2025bnf}. In particular, there exist two classes of solutions: $p$-parity solutions which are odd under the antipodal map $\Upsilon_\mathcal{H}$ and are convergent for $\tau \rightarrow \pm\infty$, and the $q$-parity solutions which are even under the antipodal map $\Upsilon_\mathcal{H}$ and are divergent for $\tau \rightarrow \pm\infty$. Since $I_{\tau A}$ falls off as $e^{-2\vert\tau\vert}$ from direct evaluation, it only contains $p$-parity tensor harmonics which satisfy $ \Upsilon_\mathcal{H}^*I_{ab}=-I_{ab}$. Evaluating this equality for the $\tau A$ component and using Eq. \eqref{IetaabtauA+} we prove Eq. \eqref{idDA}.

Now $J_{ab}$ is an SDT tensor that contains a homogeneous piece and an inhomogeneous piece. By direct evaluation, we observe that the source $S_{ab}$ is subleading by at least an order $e^{-2\vert\tau\vert}$ as compared to the leading behavior of $J_{ab}$.
We show in Appendix \ref{sec:asymptBehav} that such a source does however affect the leading behavior of $J_{ab}$. More specifically, only the $p$-parity source $S^{(p)}_{ab}=2I_{ab}$ affects this leading behavior. Since $I_{ab}$ satisfies a homogeneous equation on the hyperboloid, it will be entirely determined by its asymptotic behavior. We conclude that the contribution from the source in the leading behavior of $J_{ab}$ only depends on $\mathcal D_A^{-(0)}$ and originates from a solution to $(\square-2)^2J_{ab}=0$. Regarding the homogeneous piece, the leading behavior of $J_{ab}$ corresponds to the $q$-parity (even) piece of $J_{ab}$. Except for Eq. \eqref{idCAB} which will be proven after Eq. \eqref{map1}, the main results have been derived.  


\emph{Antipodal maps as conservation and flux balance laws on $dS_3$.} By asymptotic matching conditions, the corners $\scri^+_-$ and $\scri^-_+$ are equivalent, respectively, to the future $i^+_0$ and past $i^-_0$ boundaries of spatial infinity. Antipodal relationships between scalar, vector or tensor fields on $dS_3$ defined at the corners are generically equivalent to conservation laws of charges across spatial infinity \cite{Compere:2025bnf}. The equivalence is realized as follows. 

At leading order at spatial infinity, we distinguish two physical fields: $\sigma$ and $k^{(B)}_{ab}$ (the field $\Phi$ corresponds to the supertranslation frame). The conserved supertranslation charges at spatial infinity built from $\sigma$ are equal to the supertranslation  charges built from the Bondi mass aspect evaluated at the corners \cite{Troessaert:2017jcm,Prabhu:2019fsp,Magdy:2021rmi,Compere:2023qoa}. 

Regarding $k_{ab}^{(B)}$, we consider a $q$-parity $(B)$ type SDT tensor $\chi_{ab}$ as defined in \cite{Compere:2025bnf} that obeys $(\Box-3)\chi_{ab}=0$. Such a tensor admits a one-to-one and local map to a $(B)$ type tensor over the sphere $Y_{AB}^{(B)}:=\epsilon_{C(A}\nabla_{B)}\nabla^{C} \tilde T$ :
\begin{equation}
    \chi_{AB}(\tau\to+\infty,x^A)=-\frac{1}{8}Y_{AB}^{(B)}e^{\tau}(\tau+\mathcal{O}(\tau^0))+o(e^{0 \tau}). 
\end{equation}
Hence, the charge 
\begin{equation}
   Q^T_\chi[k_{ab}]:=  \int_{S^2} \frac{d^2x}{4 \pi} \sqrt{-q} n_a \bigg(\chi_{bc} \mathcal{D}^a k^{bc}- k^{bc}\mathcal{D}^a \chi_{bc} \bigg)\label{kabcharges}
\end{equation}
is conserved at spatial infinity and its value at any $\tau$ is given by
\begin{align}\label{CBCons}
& \int_{S^2} \! \frac{d^2 \Omega}{16 \pi} \epsilon_{C(A}\!\nabla_{B)}\!\nabla^{C} \tilde T\,   C_{AB}^{+ (0)} = \int_{S^2}   \frac{d^2 \Omega}{4 \pi} \tilde T \tilde {\mathcal M}^{+(0)}
\end{align}
where $\tilde {\mathcal M}^{+(0)}:=\frac{1}{4}\epsilon^{AB}\nabla_A \nabla^C C_{CB}^{+(0)}$, which is nothing else than the dual supertranslation charges \cite{Kol_2019,Godazgar_2019N}. Given that the electric part of $k_{ab}$ is an $n=-1$ SDT harmonic in the sense of \cite{Compere:2025bnf}, the charge \eqref{kabcharges} vanishes for any electric-type $\chi_{ab}$. Moreover, the charge \eqref{kabcharges} vanishes when $\chi_{ab}$ is a magnetic $p$-parity harmonic since $k^{(B)}_{ab}$ admits no $q$-parity harmonics. We therefore uniquely identified the non-vanishing tensorial charge associated with $k_{ab}$.

At subleading order at spatial infinity, we distinguish three physical fields: the $q$-parity part of $J_{ab}$, $I_{ab}$ and the $p$-parity part of $J_{ab}$ (which will be discussed elsewhere). There is a bijection between SDT tensors and transverse vectors on $dS_3$ \cite{Compere:2025bnf}. We start by defining a $p$-parity vector $\chi^{(p)a}_Y$ and a $q$-parity vector $\chi^{(q)a}_Y$ obeying $(\Box-1)\chi^{(p)a}_Y=(\Box-1)\chi^{(q)a}_Y=0$. Such vectors obey $\Upsilon^*_{\mathcal{H}}\chi_Y^{(p)a}=\chi_Y^{(p)a}$ and $\Upsilon^*_{\mathcal{H}}\chi_Y^{(q)a}=-\chi_Y^{(q)a}$. There is a bijection between such vectors and vectors over the sphere $Y^A$ given by 
\begin{align}
\chi^{(q)A}_Y(\tau\to+\infty,x^A) =- Y^A(x^A) e^{-\tau}+o(e^{-\tau}), \\ 
\chi_Y^{(p)A}(\tau\to+\infty,x^A) = 4 Y^A(x^A) e^{-3\tau}+o(e^{-3\tau}). 
\end{align}
Then the charge
\begin{equation}
   Q^V_\chi[V_{a}]:=  \int_{S^2} \frac{d^2x}{8 \pi} \sqrt{-q} n_a \bigg(\chi^{b}\mathcal{D}^a V_{b}-V_{b} \mathcal{D}^a \chi^{b} \bigg)\label{Vacharges}
\end{equation}
is conserved if $(\Box-1)V^a=0$ and is non-vanishing only if $V^a$ and $\chi^a$ have opposite $p$/$q$ parities \cite{Compere:2025bnf}. 

The charge $Q^V_{\chi^{(q)}}[I_a]$ for $I_a := \cosh \tau n^b I_{ab}$ is therefore conserved. Its value is given by 
\begin{align}
& \int_{S^2} \frac{d^2\Omega}{8\pi} Y^{A}  \mathcal D^{+(0)}_{A}= -\int_{S^2} \frac{d^2\Omega}{8\pi}   (\Upsilon^* Y^{A} ) \mathcal D^{-(0)}_{A}. 
\end{align}
This is a new conserved charge encoding features of the peeling property defined at spatial infinity. 

Now, the charge $Q^V_{\chi^{(p)}}[J_a]$ for $J_a := \cosh \tau n^b J_{ab}$ is finite but not conserved because $J_{ab}$ obeys an inhomogeneous equation. Denoting $S_a := \cosh \tau n^b S_{ab}$, the equation of motion reads as $(\square -1)J_a=S_a$. The current $\mathcal J^a := \chi^{(p)b}_Y \mathcal D^a J_b - J_b \mathcal D^a \chi^{(p)b}_Y$ obeys $\mathcal D_a \mathcal J^a = \chi^{(p)a}_Y S_a$.
The charge $Q^V_{\chi^{(p)}}[J_a]$ therefore obeys the following flux-balance law across spatial infinity
\begin{align}
 &\int_{S^2} \frac{d^2\Omega}{8\pi}  Y^A \nabla^B\mathcal{C}^{+(1)(0)}_{AB}  -  \int_{S^2} \frac{d^2\Omega}{8\pi}(\Upsilon^* Y^A)  \nabla^B\mathcal{C}^{-(1)(0)}_{AB}\nonumber \\ 
 & =\int_{\text{dS}_3} \frac{d^3\phi}{8\pi} \sqrt{-q} \, \chi_Y^{(p)a} S_a. \label{map1}
\end{align}
As described in Appendix \ref{sec:asymptBehav}, the inhomogeneous terms fall off sufficiently fast as $\tau \rightarrow \pm \infty$ such that the source integral converges. Given the antipodal map of $C^{\pm(B)(0)}_{AB}$ proven in Eq. \eqref{CBCons}, we can replace $\mathcal{C}^{\pm(1)(0)}_{AB}$ by $C^{\pm(1)}_{AB}$ in Eq. \eqref{map1}. Given that only the $\Upsilon_{\mathcal H}$-parity even
part ($p$-parity) of the vector source contributes, see Eq. \eqref{SourceDiff}, only $S_a^{(p)} := 2 I_a$ contributes. This equation defines the source term of Eq. \eqref{idCAB} as $\mathcal S_A[\mathcal D^{-(0)}](x^B) := \int_{\text{dS}_3} d^3\phi \sqrt{-q} \mathcal P_A^a (\phi^b , x^B) I_a(\phi^b)$ where we used the boundary-to-bulk map to define $ \mathcal P^a_A (\phi^b , x^A)$ through $\chi_Y^{(p)a}(\phi^b)=\int_{S^2}d^2\Omega\, \mathcal P_A^a (\phi^b , x^B)Y^A(x^B)$. Localizing the integral \eqref{map1} by replacing $Y^A(x^B)$ by a distribution and applying the antipodal map $\Upsilon$ proves Eq. \eqref{idCAB}.

For $n$-body scattering, the leading magnetic piece of the shear vanishes and $\nabla^B\mathcal{C}^{\pm(1)(0)}_{AB}=\nabla^BC^{\pm(1)}_{AB}$ \cite{Strominger:2014pwa}. The latter quantity is proportional to the leading behavior as $u^\pm\to-\infty$ of $(u^\pm)^2 \mathcal{J}_A$, where $\mathcal{J}_A$  appears in the Weyl scalar $\Psi_3 = r^{-1}\mathcal{J}_A \bar m^A+O(r^{-3})$ in the Bondi tetrad for vacuum Einstein's equations \cite{Geiller:2024bgf,Geiller:2024ryw}. 

\emph{Discussion.} We proved three identities \eqref{MAIN} relating the late advanced limit of $\scri^-$ and the early retarded limit of $\scri^+$ under two main hypotheses: (i) the existence of polyhomogeneous Bondi expansions at $\scri^\pm$, and (ii) the asymptotic boundary conditions on the shear \eqref{CABasymptoticsscri} in the approach to spatial infinity.  We also reformulated these identities in terms of conservation and flux balance laws of asymptotic charges at spatial infinity. 

Let us first discuss the significance of our identities \eqref{MAIN} in the absence of incoming radiation and in the absence of matter. The absence of energy flux at $\scri^-$ is equivalent to $N_{AB}^- = 0$, which implies in particular the absence of tails  $C^{(1)-}_{AB}  = 0$. Our identity \eqref{idCAB} then fixes the outgoing tails $C^{(1)+}_{AB}$ in terms of the initial field $\mathcal D^{-(0)}_A$. The Bondi mass aspect is conserved on $\scri^-$, $\partial_v m^- = 0$, as a consequence of the analogous Eq. \eqref{FluxBalancedLawsm} at $\scri^-$. The Bondi mass aspect $m^-$ is therefore defined from its value at past timelike infinity $i^-$. In general gravitational scattering, we have 
\begin{align}
m^{-}(x^A)&= \sum_{a=1}^{n^-} \frac{m_a}{\gamma_a^3 (1+v^a_i n_i(x^A))^3}, \\
C^{\pm (B)(0)}_{AB}&=0,
\end{align}
where the sum is over all $n^-$ incoming massive bodies of masses $m_a$ and velocities $v_a$ defined at $i^-$,  $m_a := E_a/\gamma_a$, $v^a_i:=P^a_i/E_a$ and $\gamma_a := 1/\sqrt{1-v^a_i v^a_i}$. We can check that for one incoming and one outgoing body, using the fact that $C^{+(1)}_{AB}=0$ \cite{Sahoo:2018lxl}, the identity \eqref{idDA} reduces to 
\begin{align}
   \Upsilon^*\nabla^B D^{+(0)}_{AB} = -\nabla^B D^{-(0)}_{AB},  
\end{align}
which is consistent with peeling that requires $D^{\pm (0)}_{AB}=0$. For more generic $n$-body scattering that may admit incoming radiation, the peeling property is broken generically as $D_{AB}^{\pm (0)}$ do not vanish, except by finely tuning the leading tail with respect to the Bondi mass aspect. Peeling is also generically broken for spacetimes admitting a non-zero magnetic piece $C^{\pm(B)(0)}_{AB}$.

Conversely, if one imposes  $D_{AB}^\pm=0$ and no incoming radiation (which implies $C_{AB}^{(1)-}=0$), then $C^{(1)+}_{AB}$ is fixed from the identity \eqref{idCAB}. Using $m^{+(0)}=\Upsilon ^* m^{-(0)}$ and Eq. \eqref{idCABB}, Eq. \eqref{idDA} reduces to 
\begin{equation}\label{eq5}
       3 P_i m^{-(0)} \nabla_A n_i +  (E + P_i n_i) \nabla_A m^{-(0)}=0. 
\end{equation}
and 
\begin{equation}\label{eq6}
       3 P_i \mathcal{\tilde M}^{-(0)} \epsilon_{AB}\nabla^B n_i +  (E + P_i n_i)\epsilon_{AB} \nabla^B \mathcal{\tilde M}^{-(0)}=0. 
\end{equation}
The unique solution to these equations is exactly the Bondi mass of a single massive body $m^{-(0)}=\frac{m}{\gamma^3 (1+ v_i n_i)^3}$ and $\mathcal{\tilde M}^{-(0)}=0$ (we used the fact that $\mathcal{\tilde M}^{-(0)}$ does not contain $\ell=0,1$ harmonics). This proves that the only asymptotically simple solutions to Einstein's equations with no incoming radiation obeying our boundary conditions admit at most one incoming and one outgoing massive body. It also shows that such solutions admit a vanishing leading-magnetic shear $C^{\pm(B)(0)}_{AB}$. Note that there are also finely-tuned cases where Eqs. \eqref{eq5}-\eqref{eq6} hold even in the presence of incoming radiation as long as $\nabla^B D_{AB}^{\pm(0)} = \frac{1}{2} \nabla_C \nabla_{\langle A} \nabla_{B\rangle} C^{\pm (1) BC}$.

We end with three final comments. First, the antipodal map \eqref{idCAB} shows that it is  consistent to simultaneously assume no incoming radiation ($N_{AB}^-=0$) and a non-vanishing tail $C^{+(1)}_{AB}$, as was previously heuristically argued by Christodoulou in \cite{christodoulou2002global} (see reasoning between Eqs. (8) and (11) and see Sec. 1.2. of \cite{Kehrberger:2021uvf} for a review). 
Second, we proved that the field $C^{+(1)}_{AB}$ which is generically non-vanishing in the analysis of  \cite{Saha:2019tub,Choi:2024ajz} can be determined in terms of the initial scattering state from Eq. \eqref{idCAB}. Third, we expect that the second and third identities \eqref{idCAB}-\eqref{idDA} can be used to formally prove the classical logarithmic soft graviton theorem \cite{Cachazo:2014fwa,Laddha:2018myi,Laddha:2018vbn} and the soft contributions to the subleading soft graviton theorem \cite{Donnay:2020lur,Freidel:2021dfs,Agrawal:2023zea,Choi:2024ajz,Geiller:2024ryw}.

\emph{Acknowledgments.}  G.C. is Research Director of the F.R.S.-FNRS. We acknowledge the use of the xAct and the RGTC packages for Mathematica. 

%


\section*{Supplementary Material}
\label{app:content}
\appendix
\section{Einstein equations at future null infinity}
\label{sec:EinsteinNullInfty}

We provide here the solution to Einstein's equations at future null infinity in Bondi gauge for a polyhomogeneous expansion of the form \eqref{BondihExp}. The computations in this Appendix generalize \cite{Geiller:2024bgf,Madler_2016,Geiller:2024ryw,Flanagan:2015pxa}.

We work at second order in the Bondi-Sachs expansion using the ansatz \eqref{BondihExp}. Stress-energy tensors compatible with the expansion \eqref{BondihExp} take the form  
\begin{align}
    T_{uu}=& \frac{T^{(0)}_{uu}}{r^2}+\frac{T^{(1)}_{uu}}{r^3}+o(r^{-3}),\label{Tuu}\\
    T_{ur}=& \frac{T^{(0)}_{ur}}{r^3}+o(r^{-3}),\\
     T_{rr}=&\frac{T^{(0)}_{rr}}{r^3}+\frac{T^{(1)}_{rr}}{r^4}+o(r^{-4}),\\
    T_{uA}=&\frac{T^{(0)}_{uA}}{r}+\frac{T^{(1)}_{uA}+T^{(1,log)}_{uA}\log r}{r^2}+o(r^{-2}),\\
     T_{rA}=&\frac{T^{(0)}_{rA}}{r^2}+\frac{T^{(1)}_{rA}}{r^3}+o(r^{-3}),\\     T_{AB}=&T^{(0)}\gamma_{AB}+\frac{T^{(1)}_{\langle AB\rangle}+T^{(1)}\gamma_{AB}}{r}+o(r^{-1})\label{TAB},
\end{align}
where all coefficients are  $u$-dependent tensor fields over the sphere.

We impose that the stress-energy tensor is divergence-free with respect to the four-dimensional metric. This implies that the coefficients in the expansion satisfy the identities  
\begin{align}
    &T_{ur}^{(0)} = - \nabla^A T^{(0)}_{rA}-\frac{1}{2} \nabla^2 T^{(0)}_{rr},\\
    &T^{(0)}= -\frac{1}{2} \partial_uT_{rr}^{(0)},\\
    &T^{(0)}_{uA}=-\partial_u T^{(0)}_{rA}-\frac{1}{2}  \nabla_A \partial_uT^{(0)}_{rr},\\
    &T^{(1)}= -\frac{1}{4} (\nabla^2+2)T_{rr}^{(0)}+ T_{rr}^{(0)}\partial_u\beta_1-\frac{1}{2}\partial_uT_{rr}^{(1)} ,\\
    &T^{(1,log)}_{uA}= \frac{1}{2} \nabla^B (\partial_u T_{rr}^{(0)}C_{AB})-\frac{1}{4} \nabla_A (\nabla^2 +2)T_{rr}^{(0)}\nn\\
    &\qquad\qquad+\nabla^B T^{(1)}_{\langle A B\rangle } -\partial_u T_{rA}^{(1)}-\frac{1}{2}\nabla_A \partial_u T_{rr}^{(1)}.
\end{align}
Solving the hypersurface equations $G_{r\mu}=8\pi T_{r\mu}$ yields the following algebraic relations: 
\begin{subequations}
\begin{align}
    \beta_1&=-2 \pi T_{rr}^{(0)},\\
    \beta_2&= -\frac{1}{32}C_{AB}C^{AB}-\pi T_{rr}^{(1)},\\
    U_2^A&= -\frac{1}{2} \nabla_B C^{AB}-6 \pi \nabla^A  T_{rr}^{(0)}-8\pi T_{rB}^{(0)}\gamma^{AB},\\
    U_{3,1}^A &=-\frac{2}{3}\nabla_B D^{AB}-\frac{8\pi}{3} \nabla^AT_{rr}^{(1)} -\frac{16\pi}{3}T_{rB}^{(1)}\gamma^{AB}.
\end{align}    
\end{subequations}
The remaining Einstein equations lead to the following evolution equations:
\begin{align}
      \partial_u m &= -\frac{1}{8} N_{AB}N^{AB} + \frac{1}{4} \nabla_A \nabla_B N^{AB}\nn\\&+3 \pi (\nabla^2+\frac{2}{3})\partial_uT_{rr}^{(0)}+4\pi \nabla^A \partial_uT_{rA}^{(0)}\nn\\&-4 \pi T_{uu}^{(0)}\label{FluxBalancedLawsm},\\
      \partial_u \mathcal{P}_A &=\nabla^B \mathcal{M}_{AB} + \frac{1}{2} C_{AB} \nabla_C N^{BC}\nn\\&-48\pi^2 \partial_uT_{rr}^{(0)}\nabla_AT_{rr}^{(0)}-24 \pi^2 T_{rr}^{(0)}\partial_u\nabla_AT_{rr}^{(0)}\nn\\&-64 \pi^2 \partial_u T^{(0)}_{rr} T^{(0)}_{rA}-32\pi^2 T^{(0)}_{rr}\partial_u  T^{(0)}_{rA}\nn\\&-4\pi C_{AB}\partial_u\nabla^B T_{rr}^{(0)}-6 \pi N_{AB}\nabla^B T_{rr}^{(0)}\nn\\&-2 \pi \partial_u(T_{rr}^{(0)} \nabla^B  C_{AB}) - 8\pi \partial_u (C_{A}^B T_{rB}^{(0)})\nn\\&-\pi \nabla_A(\nabla^2+2)T_{rr}^{(0)}+4\pi \nabla_A \nabla^B T_{rB}^{(0)}\nn\\&-4\pi (\nabla^2-1) T_{rA}^{(0)}+\frac{8}{3} \pi \partial_u T_{rA}^{(1)}-8\pi T_{uA}^{(1)}\nn\\&+\frac{7}{3}\pi \partial_u \nabla_A T_{rr}^{(1)} +4\pi \nabla^B T_{\langle A B \rangle}^{(1)}\label{FluxBalancedLawsN},\\
      \partial_ u D_{AB} &=-8\pi T_{\langle AB \rangle }^{(1)}-4\pi C_{AB} \partial_u T^{(0)}_{rr}\nn\\&+4\pi \nabla_{\langle A}\nabla_{B\rangle}T^{(0)}_{rr}\label{FluxBalancedLawsD},
\end{align}
where we defined the covariant mass tensor $\mathcal{M}_{AB}$ and the covariant Bondi angular momentum aspect $\mathcal P_A$ as
\begin{align}
    \mathcal{M}_{AB}&:=m \gamma_{AB}+\frac{1}{16} \partial_u \left(C_{CD}C^{CD}\right)\gamma_{AB} \nn\\&+ \frac{1}{2} \nabla_{[A} \nabla^C C_{B]C} - \frac{1}{8} \epsilon^{CD} C_{DE} N^{E}_C \epsilon_{AB}\nn\\&=\mathcal{M}\gamma_{AB}+\tilde{\mathcal{M}}\epsilon_{AB},\\
    \mathcal P_A &:=\frac{3}{32} \partial_A\left(C_{B C} C^{B C}\right)-\frac{4}{3}\nabla^BD_{AB}\nn\\&-\frac{3}{2}\Big(U_{3\,A}-\frac{1}{2} C_{AB} \nabla_C C^{B C}\Big).
\end{align}
Here $\mathcal{M}:=m+\frac{1}{16} \partial_u \left(C_{CD}C^{CD}\right)$ is the covariant mass and $\tilde{\mathcal{M}}:=\frac{1}{4} \nabla_C \nabla^E\epsilon^{CD} C_{DE}- \frac{1}{8} \epsilon^{CD} C_{DE} N^{EC}$ is the covariant dual mass. They correspond, respectively, to the real and imaginary parts of the $1/r^3$ component of the Weyl scalar $\Psi_2$ for vacuum Einstein's equations (see (2.13) in \cite{Geiller:2024ryw}). 
 The covariant Bondi angular momentum aspect corresponds to the $1/r^4$ component of the Weyl scalar $\Psi_1$ for vacuum Einstein's equations (see (2.13) in \cite{Geiller:2024ryw}).

\section{Einstein equations at spatial infinity}
\label{sec:EIN}

At spatial infinity, Einstein's equations for metrics admitting a Beig-Schmidt expansion \eqref{BSExpansion} reduce to a set of equations on $dS_3$ for each field appearing in the expansion. Since we do not assume any stress-energy tensor contribution at $i^0$, Einstein's equations near spatial infinity are simply given by $G_{\mu\nu}=0$. The resulting equations were derived in Appendix C of \cite{Compere:2011ve}. 

At first order in the Beig-Schmidt expansion we have 
\begin{equation}
    (\Box + 3)\sigma = 0,~~~~~ \mathcal{D}^b k_{ab} = 0,~~~~~(\Box - 3)k_{ab} = 0,
\label{BSEquationOrder1}
\end{equation}
where we used $k := q^{ab}k_{ab}=0$. These equations are linear and homogeneous and $k_{ab}$ is a symmetric divergence-free and traceless (SDT) tensor. 

At second order, there are two sets of equations that  correspond, respectively, to the logarithmic and non-logarithmic branches. First, the equations for $i_{ab}$ read as 
\begin{equation}
   i_a^a = 0,~~~~~ \mathcal{D}^b i_{ab} = 0,~~~~~(\Box - 2)i_{ab} = 0\label{iabequation}.
\end{equation}
These equations are linear and homogeneous. They are solved in \cite{Compere:2025bnf}. The equations for $j_{ab}$ are non-homogeneous:
\begin{align}
  &j_a^a= 12 \sigma^2+\frac{1}{4} k^{ij} k_{ij}+k_{ij} \mathcal{D}^{i}\mathcal{D}^{j}\sigma+\mathcal{D}_{i}\sigma \mathcal{D}^{i}\sigma\label{jabTrace},\\
  &\mathcal{D}^b j_{ab}=16 \sigma \mathcal{D}_a\sigma+2 \mathcal{D}^i\sigma \mathcal{D}_i\mathcal{D}_a\sigma+ \mathcal{D}^i\mathcal{D}^j\sigma \mathcal{D}_a k_{ij}\nn\\&\qquad\qquad+\frac{1}{2} k^{ij} \left(\mathcal{D}_i k_{aj}-\frac{1}{2} \mathcal{D}_a k_{ij}+2 \mathcal{D}_a\mathcal{D}_i\mathcal{D}_j\sigma\right)\label{jabDivergence} ,\\
  &(\Box - 2)j_{ab} = 2 i_{ab} + NL_{ab}(\sigma,\sigma)+NL_{ab}(\sigma,k)\nn\\&\qquad\qquad\qquad+NL_{ab}(k,k),\label{jabEOM}
\end{align}
where the non-linear terms $NL_{ab}(\sigma,\sigma)$, $NL_{ab}(\sigma,k)$ and $NL_{ab}(k,k)$ are  
\begin{align}
    NL_{ab}(\sigma,\sigma) :=&q_{ab} \left(6 \mathcal{D}_i\sigma \mathcal{D}^i\sigma-18 \sigma^2\right)+14 \sigma \mathcal{D}_a\mathcal{D}_b\sigma\nn\\&+8 \mathcal{D}_a\sigma \mathcal{D}_b\sigma+2\mathcal{D}^i\sigma \mathcal{D}_i\mathcal{D}_a\mathcal{D}_b\sigma\nn\\&+2\mathcal{D}^i\mathcal{D}_a\sigma \mathcal{D}_i\mathcal{D}_b\sigma,\\
    NL_{ab}(k,k) :=&k_a^i k_{ib}+k^{ij} (\mathcal{D}_i\mathcal{D}_j k_{ab} - \mathcal{D}_i\mathcal{D}_{(a} k_{b)j})\nn\\&-\frac{1}{2} \mathcal{D}_a k^{ij} \mathcal{D}_b k_{ij}+\mathcal{D}^i k_{(a}^j \mathcal{D}_{b)} k_{ij}\nn\\& -\mathcal{D}^i k_a^j \mathcal{D}_j k_{bi}+\mathcal{D}^i k_a^j\mathcal{D}_i k_{bj},\\
    NL_{ab}(\sigma,k) :=& 4 \sigma k_{ab} -2 q_{ab} k^{ij} \mathcal{D}_i\mathcal{D}_j\sigma +4 k^i_{(a} \mathcal{D}^{}_{b)} \mathcal{D}_i\sigma\nn\\&+4 \mathcal{D}^i\sigma (\mathcal{D}_{(a} k_{b)i} - \mathcal{D}_i k_{ab})\nn\\&+ k^{ij}  \mathcal{D}_{(a} \mathcal{D}_{b)} \mathcal{D}_i \mathcal{D}_j\sigma +2 \mathcal{D}_{(a} k^{ij} \mathcal{D}_{b)} \mathcal{D}_i \mathcal{D}_j\sigma  \nn\\&+\mathcal{D}^i \mathcal{D}^j\sigma  \mathcal{D}_{(a} \mathcal{D}_{b)} k_{ij}.\label{NLsigmakab}
\end{align}
\section{Residual transformations at spatial infinity}
\label{sec:residualTrans}
 One can show that the expansion \eqref{BSExpansion} is preserved under Lorentz transformations, logarithmic translations and supertranslations, which are the residual transformations preserving metrics of the form \eqref{BSExpansion} \cite{beig1982einstein,Compere:2011ve}. In this section, we give the explicit transformations of the fields under logarithmic translations and supertranslations
 and construct second order SDT tensor fields that are invariant under those transformations. We will not detail the Lorentz transformations here. 
 
 The logarithmic translations are generated by a scalar $H$ that obeys $\mathcal D_a \mathcal D_b H + q_{ab} H = 0$. Logarithmic translations act on the fields as \cite{Compere:2023qoa}\footnote{We note a typo in Eq. (212) of \cite{Compere:2023qoa}: the last term should be proportional to $5/2$ instead of $3/2$.} : 
\begin{align}
\delta_H \sigma & = H , \label{sigmaTransH}\\
\delta_H k_{ab} &= 0, \\
\delta_H i_{ab} &=\mathcal{D}_c \left( \mathcal{D}^c H (\mathcal{D}_a \mathcal{D}_b \sigma + q_{ab} \sigma) \right)\nn\\&-(\mathcal{D}_{(a} k_{b)c}-\mathcal{D}_ck_{ab})\mathcal{D}^cH
 , \\
\delta_H j_{ab} &= 
+ 10 \sigma H q_{ab} - 4 \mathcal{D}_{(a}H \mathcal{D}_{b)}\sigma \nn\\&+ 2 H D_{a} D_{b} \sigma + 2 q_{ab} D_{c} \sigma D^{c} H 
+ \frac{1}{2} D_{c} (k_{ab} D^{c} H) \nn\\&+ \frac{5}{2}D_{c} \left( D^{c} H(D_{a} D_{b} \sigma + q_{ab} \sigma)  \right)\nn\\&-\frac{1}{2}(\mathcal{D}_{(a} k_{b)c}-\mathcal{D}_ck_{ab})\mathcal{D}^cH.
\end{align}
The Beig-Schmidt supertranslations are generated by an arbitrary scalar $\omega(\phi^a)$ which obeys $(\mathcal D_c \mathcal D^c+3) \omega = 0$. This wave equation admits two sets of solutions, one set corresponding to supertranslations on $\mathcal I^\pm$ and the other set corresponding to trivial gauge transformations \cite{Troessaert:2017jcm}. Under infinitesimal supertranslations, the Beig-Schmidt fields transform as \cite{Compere:2023qoa} : 
\begin{align}
\delta_{\omega} \sigma & =0, \label{sigmaSTransl}\\
\delta_{\omega} k_{a b} & = 2\left(\mathcal{D}_a \mathcal{D}_b+q_{ab}\right) \omega \label{ktransfo}, \\
\delta_{\omega} i_{a b} & =0, \\
\delta_{\omega} j_{a b} & = k_{c(a} \mathcal{D}_{b)} \mathcal{D}^c \omega +k_{a b} \omega \nn \\
&+\mathcal{D}^c \omega \left(\mathcal{D}_c k_{a b}-\mathcal{D}_{(a} k_{b) c}\right) -4 \sigma \omega  q_{ab}\nn \\
& +4 \mathcal{D}_{(a} \sigma \mathcal{D}_{b)} \omega+\mathcal{D}_a \mathcal{D}_b\left(-\sigma \omega +\mathcal{D}_c \sigma \mathcal{D}^c \omega \right).\label{deltajab}
\end{align}

Using Eqs. \eqref{sigmaTransH}-\eqref{deltajab} and the equation of motion in Appendix \ref{sec:EIN}, we construct the following second order invariant SDT tensors 
\begin{align}
    I_{ab} :=& i_{ab}-\mathcal{D}_c \left( \mathcal{D}^c \eta~(\mathcal{D}_a \mathcal{D}_b \sigma + q_{ab} \sigma)\right)\nn\\&+(\mathcal{D}_{(a} k_{b)c}-\mathcal{D}_ck_{ab})\mathcal{D}^c\eta\label{defIab},\\
\label{defJab}
    {J}_{ab}:=& j_{ab}-\frac{5}{2} i_{ab}+\frac{1}{2} q_{ab} \Big(-2\mathcal{D}_g\sigma \mathcal{D}^g\sigma+4 \mathcal{D}^g\sigma \mathcal{D}_g\Phi \nn\\&-8 \sigma \Phi-12 \sigma^2-2 \Phi^2\Big)+\mathcal{D}^g\sigma \mathcal{D}_g\mathcal{D}_a\mathcal{D}_b\Phi \nn\\
    &+\mathcal{D}^g\Phi \mathcal{D}_g\mathcal{D}_a\mathcal{D}_b\sigma+2 \mathcal{D}^g\mathcal{D}_{(a}\sigma \mathcal{D}_{b)}\mathcal{D}_g\Phi\nn\\&-\Phi \mathcal{D}_a\mathcal{D}_b\sigma-\mathcal{D}^g\mathcal{D}_a\Phi \mathcal{D}_g\mathcal{D}_b\Phi-\sigma \mathcal{D}_a\mathcal{D}_b\Phi \nn \\
    &+2 \mathcal{D}_a\sigma \mathcal{D}_b\sigma-2 \sigma \mathcal{D}_a\mathcal{D}_b\sigma-2 \Phi \mathcal{D}_a\mathcal{D}_b\Phi\nn\\&-4 \sigma k^{(B)}_{ab} -4 \mathcal{D}_{(a}(k^{(B)}_{b)i}\mathcal{D}^i\sigma )+ 4 \mathcal{D}^i \sigma \mathcal{D}_i k^{(B)}_{ab}\nn 
            \end{align}
    \begin{align}
    &+\mathcal{D}^i \mathcal{D}^j \sigma k^{(B)}_{ij}q_{ab}+\Phi k^{(B)}_{ab} +k^{(B)}_{i(a}\mathcal{D}^i \mathcal{D}_{b)}\Phi\nn\\&-\mathcal{D}^i\Phi \mathcal{D}_{(a}k^{(B)}_{b)i}+\mathcal{D}^i\Phi \mathcal{D}_i k^{(B)}_{ab}-4\eta k^{(B)}_{ab} \nn \\& + 2  \mathcal{D}_{(a}(k^{(B)}_{b)i}\mathcal{D}^i\eta ) -\frac{5}{2}\mathcal{D}^i(\mathcal{D}_i\eta  k^{(B)}_{ab})
    +\frac{1}{2} k^{(B)}_{ac}k^{(B)\,c}_b\nn\\& - \frac{1}{8} q_{ab} (k^{(B)}_{cd}k^{(B)\,cd}+\mathcal{D}_ik^{(B)}_{cd}\mathcal{D}^ik^{(B)\,cd}) \nn\\&- \frac{1}{8} k^{(B)\,cd} \mathcal{D}_{(a} \mathcal{D}_{b)} k^{(B)}_{cd} +\frac{3}{8} \mathcal{D}_{(a} k^{(B)\,cd} \mathcal{D}_{b)}  k^{(B)}_{cd} , 
\end{align}
where $\eta$ is the odd piece of $\sigma$ satisfying $\mathcal D_a \mathcal D_b \eta + q_{ab} \eta = 0$ and $\Phi$ is the scalar defining the electric piece of $k_{ab}$.
We shall defer the discussion about the uniqueness of $I_{ab}$ and $J_{ab}$ to our detailed analysis  \cite{Compere:2026mdj}.

\section{Matching between spatial infinity and null infinities}
\label{sec:Matching}
Here, we give the explicit form of the Beig-Schmidt fields after performing the coordinate transformation from Bondi-gauge to Beig-Schmidt gauge. We start with the asymptotic behaviors at $i^0_+$ (which match with $\scri^+_-$). 
At first order, we get
\begin{align}
    \sigma &= 2 m^{+(0)} e^{-3\tau}  + o(e^{-3\tau})\label{sigmascri+-},\\
   k_{\tau\tau} &= \left(8 \nabla^{A} \nabla^B C_{AB}^{+(0)} (\tau-\frac{3}{4})- 16 ~w^{+}(y^A) \right)  e^{-3\tau}\nn\\&+ o(e^{-3\tau}),\\
    k_{\tau A} &= 2 \nabla^B C_{AB}^{+(0)} e^{-\tau} + o(e^{-\tau}),\\
    k_{AB} & = \frac{1}{2} C_{AB}^{+(0)}  e^{\tau}+ o(e^{\tau})\label{kabscri+-},
\end{align}
where $w^{+}(y^A) $ represents the trivial supertranslation gauge at $i^0_+$. Using $k_{ab}=-2 (\mathcal{D}_a\mathcal{D}_b+q_{ab})\Phi+k^{(B)}_{ab}$ and the decomposition 
\begin{equation}
    C_{AB}^{(i)} = (-2 \nabla_A \nabla_B + \gamma_{AB}\nabla^2) C^{(i)} + \epsilon_{C(A}\nabla_{B)}\nabla^C \Psi^{(i)}\label{CABiDecomposition},
\end{equation}
we deduce the fall-offs of the scalar field $\Phi$ at $i^0_+$ :
\begin{align}
    \Phi = &\frac{1}{2} C^{+(0)} e^{\tau}-\frac{1}{2} (\nabla^2+1)C^{+(0)}e^{-\tau}\nn\\&+\left(w^{+}(y^A)+\frac{1}{2}\tau \nabla^2(\nabla^2+2)C^{+(0)}\right)e^{-3\tau}\\&+ o(e^{-3\tau})\nn.
\end{align}
Since we will work with supertranslation-invariant tensors, we (partially) fix the gauge by imposing $w^+(y^A)=0$.

The subleading (order 2) Beig-Schmidt fields take the following asymptotic form:
\begin{align}
i_{\tau \tau}  =& o\left(e^{-3 \tau}\right), \label{iabtautau}\\
i_{\tau A} =  &4 e^{-2 \tau} (N_A^{+(\log )}+\nabla_B D^{+B}_A )+o\left(e^{-2 \tau}\right) ,\label{iabtauA}\\
i_{AB}  =& o\left(e^{-\tau}\right),\label{iabval}\\
j_{\tau \tau} =&  16 m^{+(1)} e^{-2 \tau}+o\left(e^{-2 \tau}\right)\nonumber ,\\
j_{\tau A} =& \nabla^B C_{A B}^{+(1)}+o\left(e^{0 \tau}\right) \label{jtauAscri+-},\\
j_{AB} =&  \frac{1}{2}C^{+(1)}_{AB} e^{2\tau}+o\left(e^{2 \tau}\right)\label{jABscri+-} . 
\end{align}
We can similarly obtain the correspondences of the fields at $i^0_-$ (near past null infinity). The analogous matching between $\scri^-_+$ and $i^0_-$ gives the same asymptotic behavior with $\tau\to-\tau$ in the trivial supertranslation gauge $w^-(x^A)=0$. Here, $w^-(x^A)=0$ is the analogous function to $w^{+}(x^A)$ and appears at order $e^{-3\vert\tau\vert}$ in the asymptotics of $\Phi$ at $\tau\to-\infty$. $w^-(x^A)=0$ and $w^+(x^A)=0$ are different trivial supertranslation gauges but this does not affect the results since we work with supertranslation-invariant quantities. 

\section{Asymptotic behaviors of sourced fields}
\label{sec:asymptBehav}

Using the equations of motion in Appendix \ref{sec:EIN} and the definition \eqref{defJab}, we find that $J_{ab}$ obeys Eqs. \eqref{BoxJab} where $S_{ab}$ is a non-dynamical source depending upon $i_{ab}$ and the subleading fields $k_{ab}^{(B)}$, $\sigma$ and $\eta$. 

In this appendix, we prove that although the source $S_{ab}$ is asymptotically subleading relative to $J_{ab}$, i.e. compare Eq. \eqref{JetaABscri+-} with Eq. \eqref{SabASY}, its $p$-parity component contributes to the leading behavior of $J_{ab}$.  The computations crucially rely on our harmonic analysis  \cite{Compere:2025bnf}.

Equation \eqref{BoxJab} is an inhomogeneous equation for $n=0$ SDT tensors studied in \cite{Compere:2025bnf}. By explicit computation, we find that $S_{ab}$ has the following behavior near the corners:
\begin{subequations}\label{SabASY}
\begin{align}
    S_{\tau\tau}(\tau\to\pm\infty,x^A)&= S^{(0)\pm}(x^A)e^{-4\vert\tau\vert} + o(e^{-5\vert\tau\vert}),\label{sourceBehav1tau}\\
    S_{\tau A}(\tau\to\pm\infty,x^A)&= \pm S^{(0)\pm}_A(x^A)e^{-2\vert\tau\vert} + o(e^{-3\vert\tau\vert}) ,\\
    S_{AB}(\tau\to\pm\infty,x^A)&= o(e^{-\vert\tau\vert})\label{sourceBehav1AB},
\end{align}    
\end{subequations}
where $S^{(0)\pm}(x^A)$, $S^{(0)\pm}_A(x^A)$ are scalars and vectors on the sphere, respectively, which depend on corner values of fields defined at $\scri^\pm$, namely $m^{\pm(0)}$, $C^{\pm(0)}_{AB}$,  $C^{\pm(1)}_{AB}$ and $D_{AB}^{\pm(0)}$. Because $J_{ab}$ is SDT, its source must also be SDT. This leads to the following relation: 
\begin{align}
    S^{(0)\pm}(x^A) = - 4 \nabla^AS^{(0)\pm}_A(x^A).
\end{align}
The leading behavior of the source is therefore entirely determined by $S^{(0)\pm}_A(x^A)$. 

To determine whether $S_{ab}$ will contribute to the antipodal matching of the leading behavior of $J_{AB}$ (see Eq. \eqref{JetaABscri+-}), we must decompose the solution of \eqref{BoxJab} into a homogeneous piece and an inhomogeneous piece. The ambiguity in this decomposition is fixed by imposing that the inhomogeneous piece has the same parity under $\Upsilon_\mathcal{H}$ as the source, for any source.

The general solution to Eq. \eqref{BoxJab} is found using the procedure described in \cite{Compere:2025bnf}. First, we define the following scalars:
\begin{subequations}
\begin{align}
    \eta^E &= \cosh^2\tau n^a n^b J_{ab}, \\ 
    \eta^B &= \cosh\tau n^a  \text{Curl}(\cosh\tau n^bJ_{ab}),\\
    \kappa^E &= \cosh^2\tau n^a n^b S_{ab}, \\
    \kappa^B &= \cosh\tau n^a  \text{Curl}(\cosh\tau n^bS_{ab}),
\end{align}    
\end{subequations}
where $\text{Curl}(V_{a}):=\epsilon_{abc} \mathcal D^b V^c$ for any vector $V^a$. $\eta^E$ and $\kappa^E$ capture the electric part of, respectively, $J_{ab}$ and  $S_{ab}$ while $\eta^B$ and $\kappa^B$ capture their respective magnetic parts. 

Eq. \eqref{BoxJab} is equivalent to the following two scalar equations
\begin{equation}
    \Box \eta^D = \kappa^D,\qquad D=E,B ,\label{scalarSDTEquation}
\end{equation}
with the sources $\kappa^D$ ($D=E,B$) admitting the following asymptotic behaviors 
\begin{equation}
    \kappa^D(\tau\to\pm\infty,x^A) = \kappa^{D(0)\pm}(x^A) e^{-2\vert\tau\vert} + o(e^{-3\vert\tau\vert}). \label{kappaAsympt}
\end{equation}

For $D=E,B$, the solution to Eq. \eqref{scalarSDTEquation} is \cite{Compere:2025bnf} :
\begin{align}
  \eta^D &= \sum_{\ell,m} A^{(D)}_{\ell m}(\tau; \bar\tau) \psi^{(p)}_{0\ell m}(\tau,x^A) 
   +B^{(D)}_{\ell m}(\tau; \bar\tau)\psi^{(q)}_{0\ell m}(\tau,x^A) \label{ScalNHSol},
\end{align}
where $\bar\tau$ is the time corresponding to a Cauchy surface, $\psi^{(c)}_{0\ell m}$, $c=p,q$, are the $n=0$ scalar harmonics built in \cite{Compere:2025bnf}, $A^{(D)}_{\ell m}(\tau ; \bar\tau) := a^{(D)}_{\ell m} + \int_{\bar\tau}^\tau d\tau' F^{(D,q)}_{\ell m}(\tau')$, $B^{(D)}_{\ell m}(\tau ; \bar\tau) := b^{(D)}_{\ell m} -\int_{\bar\tau}^\tau d\tau' F^{(D,p)}_{\ell m}(\tau')$ with 
\begin{align}\label{ASscalar}
    F^{(D,q)}_{\ell m}(\tau)&:=   \int_{S^2(\tau)}  d^2\Omega \,\kappa^D(\tau,x^A)\overline{\psi^{(q)}_{0\ell m}(\tau,x^A)} , \\
\label{BSscalar}
 F^{(D,p)}_{\ell m}(\tau)&:=   \int_{S^2(\tau)}  d^2\Omega \,\kappa^D(\tau,x^A)\overline{\psi^{(p)}_{0\ell m}(\tau,x^A)} .
\end{align}

Substituting Eq. \eqref{kappaAsympt} into Eqs. \eqref{ASscalar}-\eqref{BSscalar} and using the asymptotics of the scalar harmonics derived in \cite{Compere:2025bnf} we get 
\begin{align}
    &F_{\ell m}^{(D,q)}(\tau\to\pm\infty) = K^{(D)\pm}_{\ell m} 
    + o(e^{-\vert \tau\vert}),\\
    &F_{\ell m}^{(D,p)}(\tau\to\pm\infty) =  O(e^{-2\vert\tau\vert}),
\end{align}
where $K^{(D)\pm}_{\ell m} := -\frac{(\pm1)^{\ell}}{4}\int_{S^2} d^2\Omega \, \kappa^{D(0)\pm}\overline{Y_{\ell m}}$. We find 
\begin{align}
A^{(D)}_{\ell m}(\tau \to \pm \infty;\bar \tau)  &=(\tau-\bar \tau)    K^{(D)\pm}_{\ell m} +A^{(D)\pm}_{\ell m}(\bar \tau) + o(1),\label{AlmAsympt} \\ 
B^{(D)}_{\ell m}(\tau \to \pm \infty;\bar \tau)  &= B^{(D)\pm}_{\ell m}(\bar \tau) + O(e^{-2 \vert \tau \vert})\label{BlmAsympt}
\end{align}
where $A^{(D)\pm}_{\ell m}(\bar \tau)= a^{(D)}_{\ell m}+\int_{\bar \tau}^{\pm \infty}d\tau ' (F^{(D,q)}(\tau')-K^{(D)\pm}_{\ell m})$ and $B^{(D)\pm}_{\ell m}(\bar \tau)= b^{(D)}_{\ell m} -\int_{\bar \tau}^{\pm \infty}d\tau ' F^{(D,p)}(\tau')$ are integration constants. This solution leads to 
\begin{align}
&B^{(D)}_{\ell m}(\tau \to + \infty;\bar \tau)   -B^{(D)}_{\ell m}(\tau \to - \infty;\bar \tau)  =\nonumber \\ 
&-\int_{-\infty}^\infty d\tau' F_{\ell m}^{(D,p)}(\tau')= -\int_{\text{dS}_3} d\tau d^2\Omega  \kappa^D(x^a) \overline{\psi^{(p)}_{0\ell m}(x^a)} . \label{SourceDiff}
\end{align}
The integral over the boundary hyperboloid selects the $p$-parity $\ell m$ harmonic of the source $\kappa^D$. Thanks to Eq. \eqref{kappaAsympt} and the asymptotic behavior of $\psi^{(p)}_{0\ell m}$, this integral converges. We conclude the following about the leading ($q$-parity) behavior of $\eta^D$. We recall $\psi_{0\ell m}^{(q)}(\tau \to +\infty,x^A)=\Upsilon^* \psi_{0\ell m}^{(q)}(\tau \to -\infty,x^A)$. Using Eq. \eqref{ScalNHSol}, the leading contributions to $\eta^D$ at $\tau\to+\infty$ and at $\tau\to-\infty$ differ by a calculable term that can be explicitly built from an integration of the source $\kappa^D$ over $dS_3$.

From the solution \eqref{ScalNHSol}, we can recover the tensor solution to Eq. \eqref{BoxJab} using the decomposition in SDT harmonics (see \cite{Compere:2025bnf}): 
\begin{align}
    J_{ab} &=  \sum_{m=-1}^1 \left(a^{(E)}_m T^{(E,p),01m}_{ab}+a^{(B)}_m T^{(B,p),01m}_{ab} \right)\nn \\& + \sum_{l\ge2, m} \Big(T^{(E)\ell m}_{ab}\Big[f=\frac{(\eta^E,\psi_{0\ell m}^{(q)})\psi_{0\ell}^{(p)}(\tau)}{\sqrt{\ell(\ell+1)(\ell+2)(\ell-1)}}\Big] \nn \\&-T^{(E)\ell m}_{ab}\Big[f=\frac{(\eta^E,\psi_{0\ell m}^{(p)})\psi_{0\ell}^{(q)}(\tau)}{\sqrt{\ell(\ell+1)(\ell+2)(\ell-1)}}\Big]\nn \\&
    -T^{(B)\ell m}_{ab}\Big[\tilde f =\frac{(\eta^B,\psi_{0\ell m}^{(q)})\psi_{0\ell}^{(p)}(\tau)}{\sqrt{\ell(\ell+1)(\ell+2)(\ell-1)}}\Big]\nn \\&+T^{(B)\ell m}_{ab}\Big[\tilde f =\frac{(\eta^B,\psi_{0\ell m}^{(p)})\psi_{0\ell}^{(q)}(\tau)}{\sqrt{\ell(\ell+1)(\ell+2)(\ell-1)}}\Big]\Big),\label{SDTNHdecomp}
\end{align}
where $a^{(E)}_m$, $a^{(B)}_m$ are constants and $T^{(D)\ell m}_{ab}[f]$ ($D=E,B$) are defined in Eqs. (272)-(277) of \cite{Compere:2025bnf}. Here, we used the Klein-Gordon inner product on the scalars, which singles out the coefficients  
\begin{align}
     (\eta^E,\psi^{(q)}_{0\ell m})_{KG} &=A^{(E)}_{\ell m}(\tau;\bar\tau),\\ 
     (\eta^E,\psi^{(p)}_{0\ell m})_{KG} &=-B^{(E)}_{\ell m}(\tau;\bar\tau),\\ 
     (\eta^B,\psi^{(q)}_{0\ell m})_{KG} &=A^{(B)}_{\ell m}(\tau;\bar\tau),\\
     (\eta^B,\psi^{(p)}_{0\ell m})_{KG} &=-B^{(B)}_{\ell m}(\tau;\bar\tau).
\end{align}

\end{document}